\documentclass[letterpaper,man,floatsintext,natbib, 12pt]{article}  

\usepackage[margin=0.7in]{geometry}
\usepackage[american]{babel}
\usepackage{amsmath}
\usepackage{graphicx}
\usepackage{xcolor, colortbl}
\usepackage{setspace}
\usepackage{threeparttable}
\usepackage{endnotes}
\usepackage{xspace}
\usepackage{afterpage}
\usepackage{mathtools}
\usepackage{soul}
\usepackage{bm}
\usepackage{dsfont}
\usepackage{amssymb}
\usepackage{lscape}
\usepackage[normalem]{ulem}
\usepackage{csquotes}
\usepackage{algorithm}
\usepackage{algpseudocode} 
\usepackage{amsthm}
\theoremstyle{remark}

\graphicspath{ {figure/} }

\newcommand{\PreserveBackslash}[1]{\let\temp=\\#1\let\\=\temp}
\newcolumntype{C}[1]{>{\PreserveBackslash\centering}p{#1}}
\newcolumntype{L}[1]{>{\PreserveBackslash\raggedright}p{#1}}
\newcolumntype{R}[1]{>{\PreserveBackslash\raggedleft}p{#1}}

\makeatother

\RequirePackage[colorlinks,citecolor=black,urlcolor=black, linkcolor = black ]{hyperref}

\usepackage{ragged2e}
\usepackage{multirow}
\usepackage{subcaption}
\usepackage{amsmath,amssymb,xcolor}
\usepackage{pgf,tikz}
\usetikzlibrary{arrows,shapes.arrows,shapes.geometric,shapes.multipart, decorations.pathmorphing,positioning,shapes.swigs,shapes,decorations,arrows,calc,arrows.meta,fit,positioning}
\tikzset{
    -Latex,auto,node distance =1 cm and 1 cm,semithick ,
    state/.style ={ellipse, draw, minimum width = 0.7 cm},
    point/.style = {circle, draw, inner sep=0.04cm,fill,node contents={}},
    bidirected/.style={Latex-Latex,dashed},
    el/.style = {inner sep=2pt, align=left, sloped},
    styRectDef/.style = {rectangle, rounded corners, draw=black, inner xsep=6mm, inner sep=3mm}
}

\usepackage[style=apa,sortcites=true,sorting=nyt,backend=biber]{biblatex}
\DeclareLanguageMapping{american}{american-apa}
\usepackage{sectsty}
\usepackage{authblk}
\sectionfont{\fontsize{14}{15}\selectfont}

\usepackage{footmisc}
\renewcommand\footnotelayout{%
  \advance\leftskip 1cm
  \advance\rightskip 1.2cm
 } 
\usepackage[font=small, margin=0.4in]{caption}
\sectionfont{\fontsize{13}{15}\selectfont} 
\subsectionfont{\fontsize{13}{15}\selectfont} 
\usepackage{lipsum}

\makeatletter
\newcommand{\algorithmfootnote}[2][\footnotesize]{%
  \let\old@algocf@finish\@algocf@finish
  \def\@algocf@finish{\old@algocf@finish
    \leavevmode\rlap{\begin{minipage}{\linewidth}
    #1#2
    \end{minipage}}%
  }%
}

\begin{document}

\title{\large{Equality, Equity, and Causality in Fairness Research: \\ A Commentary on Cheng (2026)} \\~\\ }

\author[]{\normalsize{Youmi Suk}}

\affil[]{\small{Teachers College, Columbia University \\ ysuk@tc.columbia.edu}}

\date{\textcolor{white}{\small{}}} 
\maketitle

\abstract{This is an invited commentary on the Psychometrika focus article ``Fairness Issues and Evaluation in Psychometrics and AI/ML: What Can We Learn from Each Field?'' by Ying Cheng (2026, doi:10.1017/psy.2026.10110). Cheng offers a systematic comparison between long-standing test fairness and modern algorithmic fairness.  Her mapping of the entire testing workflow onto the AI/ML fairness paradigm, rather than only the final selection stage, is a crucial contribution to interdisciplinary fairness research. This commentary extends her discussion by examining two conceptual issues: the distinction between equality and equity, and the role of causality in fairness research. The focus article and this commentary point to directions for future fairness research across the psychometrics and AI/ML communities.
\\

}

\vspace{1cm}

\setcounter{secnumdepth}{3}
\textcite{cheng2026fairness} systematically compares test fairness, which emerged in the 1960s, and modern algorithmic fairness, which has expanded rapidly since the 2010s, in terms of fairness definitions, criteria, and bias mitigation. Her framework casts the testing process as a chain of actions and data, where bias can propagate as in AI/ML pipelines. The focus article also offers a to-do list for cross-disciplinary fairness research across the psychometrics and AI/ML fields, spanning technical, legal, and ethical dimensions.
 
I strongly agree with the author's call to draw parallels across the entire workflow of testing and AI/ML applications, rather than focusing only on the final selection stage; this comparison is more holistic and far-reaching than prior work on the topic. Her emphasis on holistic fairness evaluation and nuanced fairness definitions echoes Chapter 3, ``Fairness in Testing,'' of the \emph{2014 Standards} \parencite{aera2014standards}, which requires that fairness be considered throughout all stages of test development, administration, scoring, interpretation, and use. The chapter also articulates four common views of fairness: fair treatment during the testing process, the lack of measurement bias, access to the constructs measured, and validity of test score interpretations.
 
Additionally, \textcite{cheng2026fairness} identifies areas in need of fairness research on technical advances, laws and policies governing AI/ML use, and the ethics and values behind fairness, concluding with four paths for integrating AI/ML methods with psychometric methods. While I applaud the focus article's overall contribution, I offer two comments on conceptual issues that received less attention: (i) the tension between fairness as equality and fairness as equity, and (ii) the role of causality in evaluating fairness.
 
\section{Fairness: Equality versus Equity}\label{sec:def}
 
\textcite{cheng2026fairness} focuses primarily on fairness operationalized as variations of \emph{equality}, emphasizing metrics designed to achieve (conditional) parity across the testing and AI/ML communities. For example, standard group fairness criteria in AI/ML (e.g., independence, separation, and sufficiency) seek to equalize particular statistical properties across pre-defined groups, and the focus article carefully traces their counterparts in the testing literature. While this focus reflects the dominant discourse in fairness research, it does not fully engage with how evolving concepts of fairness in the testing field navigate the distinction between \emph{equality} and \emph{equity}. Notably, the testing community has conceptualized fairness from treating everyone the same (i.e., aiming for equality) to treating individuals differently according to their unique needs or characteristics (i.e., aiming for equity); the former view is embodied in standardized testing, and the latter in personalized assessment \parencite{bennett2026fairness}. This evolving conception is also reflected in the \textit{2014 Standards} \parencite{aera2014standards}. However, it is useful to distinguish two related but distinct dimensions: whether individuals receive the same or different \emph{treatment} (e.g., items, testing conditions, interventions), and whether the resulting \emph{outcomes} are equal or unequal. Equality and equity are often contrasted along the first dimension, but, as discussed below, equity is ultimately judged along the second.
 
In the AI/ML field, individual fairness offers a parallel conception, promoting the idea that similar individuals should be treated similarly by being ``aware'' of individual attributes for equitable treatment \parencite{dwork2012fairness}. When coupled with causal reasoning, individual fairness can be further refined so that two individuals with the same characteristics are treated equally in a way that is also counterfactually fair \parencite{kusner2017counterfactual}; see Section \ref{sec:cf} for more details on causal fairness. This individualized conception is closely related to personalization (i.e., whether treatments are individualized). Personalization operates on the treatment dimension, and it can serve as a means toward \emph{equity}. In algorithmic decision-making contexts, models are often designed to provide personalized interventions or policies tailored to individuals' needs and characteristics in order to maximize a target outcome (e.g., achievement scores, college enrollment). Such methods include dynamic treatment regimes, reinforcement learning, and optimal stochastic control. These methods conceptualize individual fairness through personalization, and they can also incorporate additional constraints that prioritize equality of treatment if desired \parencite{suk2026fair}.
 
Meanwhile, personalization has long been engineered in the testing field in the form of computerized adaptive testing (CAT), which moves beyond standardized assessment \parencite{bennett2026fairness}. 
Traditional CAT matches item difficulty in real time to an examinee's estimated ability, primarily to improve measurement efficiency and precision rather than to reduce outcome disparities. For personalized assessment, psychometricians now leverage advances in AI to adjust not only item difficulty but also content, format, presentation, response mode, and other conditions in real time based on a vector of examinee characteristics (e.g., cultural background, first language, individual interests).

However, equity is not simply differential treatment; its ultimate goal lies in the outcome dimension, namely, to \textit{reduce unjust outcome gaps}. Personalization tools do not guarantee this equitable outcome unless they are explicitly designed to pursue it. Personalized recommendations in AI/ML typically aim to maximize the final expected outcome, whereas personalized assessments aim to improve the precision of ability estimates or have yet to develop a specific theory of which outcomes to optimize \parencite{bennett2026fairness}. Consequently, the impact of these adaptive methods on existing outcome gaps (e.g., decrease, increase, or no change) is not expected to follow a systematic pattern \parencite{suk2026fair}. Of course, defining the aim of equity as the reduction of outcome gaps is itself a value commitment, which \textcite{cheng2026fairness} urges the field to make explicit. Overall, fairness is intertwined with the concepts of equality and equity, and its conceptualizations have shifted over time in how the two are weighted. Evaluating whether either aim is met, however, ultimately requires understanding what drives outcome disparities---a causal question, to which I now turn.  \raggedbottom
 
\section{Causal Fairness}\label{sec:cf}

Many fairness questions are fundamentally causal: researchers and policymakers want to know whether and how a protected variable (e.g., gender, race/ethnicity, socioeconomic status (SES)) \emph{causes} unjust societal outcomes (e.g., differential performance in algorithms). Here it is helpful to distinguish between determining \emph{whether} a disparity exists and understanding \emph{why} it exists. Association-based notions, such as equalized odds and differential item functioning (DIF), are valuable for the former, that is, for detecting and characterizing disparities, but not for the latter. Although \textcite{cheng2026fairness} lists causal fairness among promising future directions, the focus article primarily addresses association-based definitions and does not develop what causal reasoning adds for the latter. A key limitation of association-based notions is that they cannot uncover the mechanisms underlying unfairness, even with unlimited data. This challenge is recognized in the \emph{2014 Standards}, stating: ``In many cases, it is not clear whether the differences (in outcomes) are due to real differences between groups in the construct being measured or to some source of bias (e.g., construct-irrelevant variance or construct underrepresentation). In most cases, it may be some combination of real differences and bias.''

\begin{figure}[!h]
\centering
    \subcaptionbox{CDIF  \label{fig:DIF}}[.3\textwidth]
    {
\begin{tikzpicture}
\tikzset{line width=1pt, outer sep=1pt,
ell/.style={draw,fill=white, inner sep=3pt,
line width=1pt},
swig vsplit={gap=5pt,
inner line width right=0.5pt}};
\node[name=A,shape=swig vsplit] at (0,0){  \nodepart{left}{$A$} \nodepart{right}{$a$} };
\node[name=X,ell,shape=circle] at (2,1.5){$\Theta$};
\node[name=Y,ell,shape=ellipse] at (4,0){$Y^{a}$};
\draw [->, line width=0.75pt] (X) to (Y);
\draw [draw=blue, ->, line width=1pt] (A) to (Y);
\end{tikzpicture}}
\hspace{0.4in}
    \subcaptionbox{CDIF and Impact \label{fig:DIF_impact}}[.3\textwidth]{
\begin{tikzpicture}
\tikzset{line width=1pt, outer sep=1pt,
ell/.style={draw,fill=white, inner sep=3pt,
line width=1pt},
swig vsplit={gap=5pt,
inner line width right=0.5pt}};
\node[name=A,shape=swig vsplit] at (0,0){  \nodepart{left}{$A$} \nodepart{right}{$a$} };
\node[name=X,shape=swig vsplit] at (2,1.5){ \nodepart{left}{$\Theta^a$} \nodepart{right}{$\theta$}};
\node[name=Y,ell,shape=ellipse] at (4,0){$Y^{a, \theta}$};
\draw [->, line width=0.75pt] (A) to (X);
\draw [draw=blue, ->, line width=1pt] (A) to (Y);
\draw [->, line width=0.75pt] (X) to (Y);
\end{tikzpicture}}
  \caption{Single-world intervention graphs for causal differential item functioning (CDIF) in the absence and presence of impact. }\label{fig:SWIG_DIFs}
\end{figure}
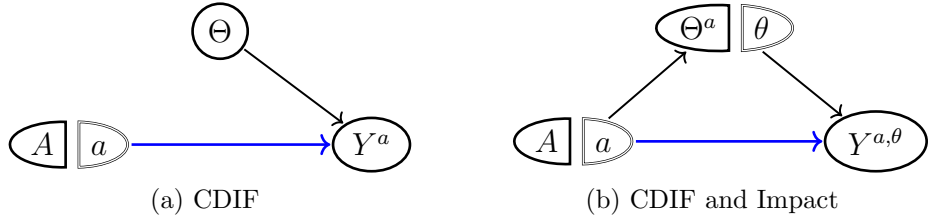

In contrast, causality-based fairness notions provide more nuanced reasoning about the mechanisms behind unfairness. Notably, in psychometrics, \textcite{suk_lyu_2026} propose a causal framework for item fairness using single-world intervention graphs (SWIGs), which combine causal graphs and potential outcomes. This approach bridges psychometric theory with algorithmic fairness and illustrates how causal reasoning can identify the specific pathways along which unfairness arises. To illustrate, Figure \ref{fig:SWIG_DIFs} presents SWIGs of causal DIF (CDIF) in the absence and presence of impact (i.e., real differences in ability across groups). Black arrows denote permissible paths, whereas blue arrows denote unfair paths. Here, $A$ represents the protected group variable (serving as a treatment), $\Theta$ represents ability (serving as a predictor in Figure \ref{fig:DIF} and a mediator in Figure \ref{fig:DIF_impact}), and $Y^{a}$ represents the potential/counterfactual item response that would be observed if an individual were assigned to $A=a$. In Figure \ref{fig:DIF_impact}, $\Theta^a$ denotes the potential ability under $A=a$, and $Y^{a,\theta}$ denotes the potential item response under $A=a$ and $\Theta^a=\theta$. For simplicity, pre-treatment confounders affecting $A$ and the potential responses are suppressed in Figure \ref{fig:SWIG_DIFs}.
 
Unlike association-based fairness, causal fairness enables researchers to raise counterfactual questions \parencite[e.g.,][]{suk_lyu_2026, kusner2017counterfactual}, such as what an individual's performance would have been had they belonged to a different group $A=a$ (e.g., low SES) instead of their original group $A=a'$ (e.g., high SES). This counterfactual scenario is visualized in Figure \ref{fig:SWIG_DIFs} by splitting the node $A$ into a random half and a fixed half, where $A$ denotes the naturally occurring (pre-intervention) treatment and $a$ denotes the value to which $A$ is set under the intervention.
 
Furthermore, causal approaches help explain why and how unfairness occurs. As shown in Figure \ref{fig:DIF_impact}, only the direct path from $a$ to $Y^{a, \theta}$ is unfair, representing CDIF, whereas the indirect path ($a \rightarrow \Theta^a$, $\theta \rightarrow Y^{a, \theta}$) is permissible because the ability $\Theta$ acts as a \emph{resolving} mediator \parencite{kilbertus2017avoiding}, i.e., an intermediate variable that provides a legitimate, construct-relevant justification for outcome differences between groups. In contrast, association-based DIF captures the overall effect, entangled with both direct and indirect effects as well as confounding bias. With causal reasoning and substantive domain knowledge, researchers can effectively distinguish genuine item bias from real latent differences (and external confounding), thereby addressing the dilemma highlighted in the \emph{2014 Standards}. Of course, these benefits are realized only when the relationships among variables are accurately specified in the causal graph and when the permissible pathways are justified by substantive and normative judgments about the construct and the intended use of the assessment. With these caveats in mind, causal inference can play a key role in promoting fairness throughout the entire testing life cycle, including test administration, scoring, and interpretation.

\section{Conclusions}\label{sec:con}
 
\textcite{cheng2026fairness} provides a systematic comparison of fairness issues (e.g., definitions, criteria, bias mitigation) between the testing and AI/ML fields, along with a discussion of areas in need of future research. To extend this discussion, I have emphasized two conceptual issues: the distinction between equality and equity, and the role of causality in fairness research. Fairness concepts have evolved in how they balance equality and equity, and they require causal reasoning to uncover the mechanisms behind unfairness. Incorporating these issues will enrich future research at the intersection of the measurement and AI/ML communities.

\printbibliography

\end{document}